# A Numerical Approach for Modeling the Shunt Damping of Thin Panels with Arrays of Separately Piezoelectric Patches

Peyman Lahe Motlagh[1], Mustafa Kemal Acar[1]
[1] Gebze Technical University, Kocaeli/Turkey, peyman.lahe@gtu.edu.tr
[1] Gebze Technical University, Kocaeli/Turkey, mkemalacar@gtu.edu.tr

*Abstract* - Two-dimensional thin plates are widely used in many aerospace and automotive applications. Among many methods for the attenuation of vibration of these mechanical structures, piezoelectric shunt damping is a promising way. It enables a compact vibration damping method without adding significant mass and volumetric occupancy. Analyzing the dynamics of these electromechanical systems requires precise modeling tools that properly consider the coupling between the piezoelectric elements and the host structure. This paper presents a methodology for separately shunted piezoelectric patches for achieving higher performance on vibration attenuation. The Rayleigh-Ritz method is used for performing the modal analysis and obtaining the frequency response functions of the electro-mechanical system. The effectiveness of the method is investigated for a broader range of frequencies, and it was shown that separately shunted piezoelectric patches are more effective.

*Keywords* - piezoelectric patches, Rayleigh-Ritz model, shunt damping, electromechanical systems

## I. Introduction

Mechanical structures are exposed to vibrations caused by operational or environmental sources, which can be undesirable, Over the decades, many passive and semi-passive systems have been proposed for reducing these vibrations [1], [2]. Piezoelectric structures have been widely used in a range of applications including vibration control [3], [4], energy harvesting [5], [6], structural health monitoring [7]. Over the past few decades, among the transducers that convert the mechanical (electrical) energy to electrical (mechanical) energy, piezoelectric transducers are mostly preferred to the electromagnetic [8], [9], and electrostatic [8], [10], ones due to their high power density and ease of manufacturing at different size scales [11]. The most common use of piezoelectric materials in the form of patches/layers is by integrating them to the surfaces of flexible beam/plate-like structures, and then utilize them in bending motion for generating an electrical signal and vice versa (applying a voltage to generate a bending deformation). Piezoelectric patches have been used for shunt damping applications that focus on designing simple electrical circuits that efficiently reduce the structural vibrations [12]. Ideally, performance requirements include stability and low energy consumption. The shunt circuit is said to be passive if it does not require an external power supply (e.g. R-shunt) and semi-passive if the circuit operation needs external power supply but does not deliver any power to the mechanical structure (e.g. SSDI) [13], [14]. A resistor connected to the piezoelectric transducer provides the simplest means of achieving energy dissipation and thus vibration damping [5]. In energy harvesting, vibration control, and actuation/sensing applications, piezoelectric materials are typically in the form of thin square patches that are bonded to specific locations on the surface of the thin plates. For the implementation of these structures, [2], [15] presents an electro-elastic model of a thin-laminated composite plate with surface bonded piezoelectric patches by considering the mass and stiffness contribution of the patches as well as the two-way electromechanical coupling effect. Based on the above studies conducted on the modeling of piezoelectric patches bonded on thin plate structures, a new approach is proposed in this study to obtain broadband frequency shunt damping. To cover the electro-mechanical equations for separately shunted circuits and neutral axis shift was included in the electro-mechanical model to accommodate when a single patch is used on one side of the host plate. In addition, an optimization study is performed to determine the optimized design parameters of the shunt circuit (i.e. Resistor values and number of patches) to minimize the vibration amplitudes at the first three modes of the structure. It was shown that the separated patches improved the shunt damping performance which was a significant observation of the present study.

## II. Analytical model of a thin plate with multiple piezoelectric patches

In this section, a brief description of the model of a thin plate with multiple piezoelectric patches is given based on the Kirchhoff plate theory [15]. Figure 1, presents the host plate with all four edges clamped (CCCC) boundary conditions and the structurally integrated piezoelectric patches in separated and connected configurations, respectively.

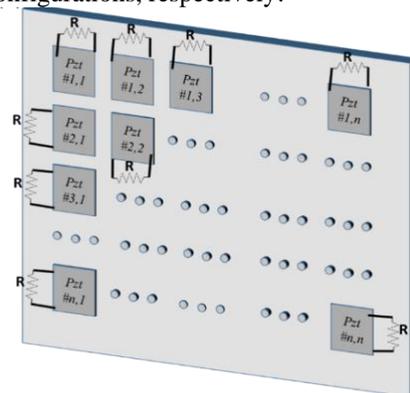

Figure 1: Separated configuration of the piezoelectric patches and the host plate

## III. CONSTITUTIVE EQUATIONS OF THE THIN PLATE WITH PIEZOELECTRIC PATCHES

Since the piezoelectric patches are typically manufactured as a thin plate, piezoelectric patch skin can be modeled as a two-dimensional Kirchhoff plate. According to Kirchhoff plate theory, the deflection of the middle surface is small compared to the thickness of the skin. Since the normal stress $\sigma_{zz}$ in the thickness direction is much smaller compared to the in-plane stresses, it can be ignored under the assumption of the thin plate theory. The material of the structural layer is assumed to be isotropic:

$$\begin{pmatrix}\sigma_{xx}\\ \sigma_{yy}\\ \tau_{xy}\end{pmatrix} = \frac{Y_s}{1-v_s^2}\begin{bmatrix}1 & v_s & 0\\ v_s & 1 & 0\\ 0 & 0 & (1-v_s)/2\end{bmatrix}\begin{pmatrix}\varepsilon_{xx}\\ \varepsilon_{yy}\\ \gamma_{xy}\end{pmatrix} \quad (1)$$

The constitutive equations of a piezoelectric patch are expressed in a reduced form as Eq. (2) [16], [17]:

$$\begin{pmatrix}\sigma_{xx}\\ \sigma_{yy}\\ \tau_{xy}\\ D_3\end{pmatrix} = \begin{bmatrix}\bar{c}_{11} & \bar{c}_{12} & 0 & -\bar{e}_{31}\\ \bar{c}_{12} & \bar{c}_{11} & 0 & -\bar{e}_{31}\\ 0 & 0 & \bar{c}_{66} & 0\\ \bar{e}_{31} & \bar{e}_{31} & 0 & \varepsilon^s_{33}\end{bmatrix}\begin{pmatrix}\varepsilon_{xx}\\ \varepsilon_{yy}\\ \gamma_{xy}\\ E_{33}\end{pmatrix} \quad (2)$$

Where $\bar{c}_{ij}$ are reduced elastic moduli of the piezoelectric patches, $\bar{e}_{ij}$ and $\varepsilon^s_{33}$ are piezoelectric constant and dielectric permittivity, respectively. Hamilton's principle is used to determine the equation of motion Eq. (3):

$$\delta \oint_{t_1}^{t_1}(KE - PE + W_p)\,dt = 0 \quad (3)$$

Where kinetic energy, potential energy, and applied external work is indicated as $KE, PE$, and $Wp$ respectively. Kinetic energy of the system can be formulated as:

$$KE = \frac{1}{2}\iint_S m(x,y)\dot{w}^2\,dS \quad (4)$$

where $S$ is the area of the system including the top surfaces of the piezo-patches and the thin composite plate. Here, $m(x,y)$ and $\dot{w}$ are the mass per unit area and the velocity terms. The equivalent mass per unit area can be derived as Eq. (5):

$$m(x,y) = \rho_s h_s + \rho_p h_p P(x,y) \quad (5)$$

Here $\rho_s, \rho_p$ are densities and $h_s$ and $h_p$ are the thicknesses of the host plate and the piezoelectric patch, respectively. The indicator function $P(x,y)$ is defined to identify the areas where $k$ piezoelectric patches are attached to the surface of the structural layer by :

$$P(x,y) = \sum_{i=1}^{k}[H(x-x_{i,1}) - H(x-x_{i,2})] \times [H(y-y_{i,1}) - H(y-y_{i,2})] \quad (6)$$

Where $x_1, x_2, y_1,$ and $y_2$ are the vertices of the area along x- and y-axes, respectively, and H denotes the Heaviside unit step function. Potential energy of the system can be written as:

$$PE = PE_s + PE_p = \frac{1}{2}\iiint_{V_s}\{(\sigma_{xx}\varepsilon_{xx})_s + (\sigma_{yy}\varepsilon_{yy})_s + (\tau_{xy}\gamma_{xy})_s\}dV_s + \frac{1}{2}\iiint_{V_p}\{(\sigma_{xx}\varepsilon_{xx})_p + (\sigma_{yy}\varepsilon_{yy})_p + (\tau_{xy}\gamma_{xy})_p\}dV_p \quad (7)$$

where, $V_s$ and $V_p$ are the volumes of the thin composite plate and the piezo-ceramic patch, respectively. The potential energy stored in the piezo-patches can be written as:

$$PE_P = \frac{1}{2}\sum_{l=1}^{2}\iint_{S_p} P(x,y)\left\{D_{11}^p\left(\frac{\partial^2 w}{\partial x^2}\right)^2 + 2D_{12}^p\left(\frac{\partial^2 w}{\partial y^2}\right)\left(\frac{\partial^2 w}{\partial x^2}\right) + D_{11}^p\left(\frac{\partial^2 w}{\partial y^2}\right)^2 + 4D_{66}^p\left(\frac{\partial^2 w}{\partial xy}\right)^2 - \bar{e}_{31}v(t)\left(\frac{h_s+h_p}{2}\right)\left(\frac{\partial^2 w}{\partial x^2} + \frac{\partial^2 w}{\partial y^2}\right)\right\}dS_p \quad (8)$$

where $S_p$ is the surface area of the piezo-patches, v(t) is the voltage and $D_{ij}^P$ is the bending stiffness matrix of the patches which can be obtained as:

$$D_{11}^p = \int_{\frac{h_s}{2}}^{\frac{h_s}{2}+h_p}\bar{c}_{11}\,z^2 dz = \int_{-\frac{h_s}{2}-h_p}^{-\frac{h_s}{2}}\bar{c}_{11}\,z^2 dz$$
$$= \bar{c}_{11}\left(\frac{h_p^3}{3} + \frac{h_s^2 h_p}{4} + \frac{h_s h_p^2}{2}\right)$$

$$D_{12}^p = \int_{\frac{h_s}{2}}^{\frac{h_s}{2}+h_p}\bar{c}_{12}\,z^2 dz = \int_{-\frac{h_s}{2}-h_p}^{-\frac{h_s}{2}}\bar{c}_{12}\,z^2 dz$$
$$= \bar{c}_{12}\left(\frac{h_p^3}{3} + \frac{h_s^2 h_p}{4} + \frac{h_s h_p^2}{2}\right)$$

$$D_{66}^p = \int_{\frac{h_s}{2}}^{\frac{h_s}{2}+h_p}\bar{c}_{66}\,z^2 dz = \int_{-\frac{h_s}{2}-h_p}^{-\frac{h_s}{2}}\bar{c}_{66}\,z^2 dz$$
$$= \bar{c}_{66}\left(\frac{h_p^3}{3} + \frac{h_s^2 h_p}{4} + \frac{h_s h_p^2}{2}\right) \quad (9)$$

The applied external work of the point force can be written as:

$$W_p = \iint_S f(t)\delta(x-x0)\delta(y-y0)\,dS \quad (10)$$

where $f(t)$ is the force amplitude, and $\delta(x)$ and $\delta(y)$ are the Dirac delta functions along the $x$ and $y$ axes.

Based on the modal expansion, the relative displacement of piezoelectric patch skin is approximated by a linear combination of the assumed modes as Eq. (11), where $\mu_{ij}(t)$ are the generalized modal coordinates, N is the total number of vibration modes in y and R is the total number of vibration modes x coordinates. Assumed modes are indicated by $U_{ij}W_{ij}(x,y)$, where $W_{ij}(x,y)$ are the trial functions satisfying the boundary conditions, and $U_{ij}$ are the corresponding coefficients.

$$w(x,y,t) = \sum_{i=1}^{N}\sum_{j=1}^{N} U_{ij}\,W_{ij}(x,y)\mu_{ij}(t) \quad (11)$$

$$[K_{rn,kl} - \omega_{rn}^2 M_{rn,kl}][U_{rn}] = \{0\} \quad (12)$$

Here, assumed mode shape coefficients $U_{ij}$'s are the eigenvectors and natural frequencies $\omega_{ij}$'s are the square root of the eigenvalues of Eq. (12). Then, the equation of motion of the plate and piezoelectric patches can be derived as Eq. (13) in which w is displacement of the system:

$$m(x,y)\ddot{w} + \left[ \{D^s + P(x,y)D^{sp}\} \left\{ \left(\frac{\partial^2 w}{\partial x^2}\right)^2 + \left(\frac{\partial^2 w}{\partial y^2}\right)^2 + 2v_s\left(\frac{\partial^2 w}{\partial x^2}\frac{\partial^2 w}{\partial y^2}\right) + 2(1-v_s)\left(\frac{\partial^2 w}{\partial x \partial y}\right)^2 \right\} + P(x,y)D_{11}^p \left\{ \left(\frac{\partial^2 w}{\partial x^2}\right)^2 + \left(\frac{\partial^2 w}{\partial x^2}\right)^2 \right\} + 2P(x,y)\left\{ D_{12}^p\left(\frac{\partial^2 w}{\partial x^2}\frac{\partial^2 w}{\partial y^2}\right) + 2D_{66}^p\left(\frac{\partial^2 w}{\partial x \partial y}\right)^2 \right\} + P(x,y)\bar{e}_{31}v(t)\left(\frac{h_s+h_p}{2}-z_0\right)\left(\frac{\partial^2 w}{\partial x^2} + \frac{\partial^2 w}{\partial y^2}\right) \right] = f(t)\delta(x-x0)\delta(y-y0) \quad (13)$$

where $D_{11}^p$, $D_{12}^p$, $D_{66}^p$, $D^s$ and $D^{sp}$ are as follows

$$D_{11}^p = \bar{c}_{11}\left(\frac{h_p^3}{3} + \frac{h_s^2 h_p}{4} + \frac{h_s h_p^2}{2} - z_0(h_p h_s + h_p^2) + z_0^2 h_p\right)$$

$$D_{12}^p = \bar{c}_{12}\left(\frac{h_p^3}{3} + \frac{h_s^2 h_p}{4} + \frac{h_s h_p^2}{2} - z_0(h_p h_s + h_p^2) + z_0^2 h_p\right)$$

$$D_{66}^p = \bar{c}_{66}\left(\frac{h_p^3}{3} + \frac{h_s^2 h_p}{4} + \frac{h_s h_p^2}{2} - z_0(h_p h_s + h_p^2) + z_0^2 h_p\right) \quad (14)$$

$$D^s = \frac{Y_s h_s^3}{12(1-v_s^2)}$$

$$D^{sp} = \frac{Y_s}{1-v_s^2}\left(\frac{h_s^3}{12} + z_0^2 h_p\right) \quad (15)$$

In the separated configuration, each patch is connected to an electric circuit independently as in Figure 1-(a), by applying Kirchhoff's current law for each patch, the circuit equation can be written as Eq. (16):

$$C_p^k\left(\frac{dv_k(t)}{dt}\right) + \frac{v_k(t)}{(Z_l)_k} = i_k(t), (k = 1,2,\dots,Number\ of\ patches) \quad (16)$$

The relationship between $W_{ij}$ and $v_k(t)$ can be written as Eq. (17):

$$\left[\frac{1}{(Z_l)_k} + j\omega C_p^k\right]V_k(t) + j\omega \sum_{n=1}^{N}\sum_{r=1}^{N}\frac{F_0 U_{ij} W_{ij}(x_0,y_0)\tilde{\theta}_{ij}^k}{\omega_{ij}^2 - \omega^2 + (2j)\xi_{ij}\omega_{ij}\omega} + j\omega \sum_{i=1}^{N}\sum_{j=1}^{N}\tilde{\theta}_{ij}^k \frac{\sum_{k=1}^{k}\tilde{\theta}_{ij}^k v_k(t)}{\omega_{ij}^2-\omega^2+(2j)\xi_{ij}\omega_{ij}\omega} = 0 \quad (17)$$

Then $v_k(t)$ are the only unknowns in this equation, and they can be derived as:

$$\begin{bmatrix}V_1\\ \vdots\\ V_k\\ \vdots\\ V_n\end{bmatrix} = A^{-1}\begin{bmatrix}b_1\\ \vdots\\ b_k\\ \vdots\\ b_n\end{bmatrix} =$$

$$\begin{bmatrix}\left\{\frac{1}{(Z_l)_1}+j\omega C_p^1+a'_{11}\right\} & \cdots & a'_{1k} & \cdots & a'_{1n}\\ \vdots & \ddots & \vdots & \ddots & \vdots\\ a'_{k1} & \ddots & \left\{\frac{1}{(Z_l)_k}+j\omega C_p^k+a'_{kk}\right\} & \ddots & a'_{kn}\\ \vdots & \ddots & \vdots & \ddots & \vdots\\ a'_{n1} & \cdots & a'_{nk} & \cdots & \left\{\frac{1}{(Z_l)_n}+j\omega C_p^n+a'_{nn}\right\}\end{bmatrix} \quad (18)$$

where $a$ and $b$'s are defined as

$$a'_{ls} = j\omega \sum_{j=1}^{N}\sum_{i=1}^{N}\frac{\tilde{\theta}_{ij}^l \tilde{\theta}_{ij}^s}{\omega_{ij}^2-\omega^2+2j\xi_{ij}\omega_{ij}\omega}, (i,j=1,\dots,N), (l,s=1,2,\dots,number\ of\ patches) \quad (19)$$

$$b_k = j\omega \sum_{i=1}^{N}\sum_{j=1}^{N}\frac{F_0 U_{ij} W_{ij}(x_0,y_0)\tilde{\theta}_{ij}^k}{\omega_{rn}^2-\omega^2+2j\xi_{rn}\omega_{rn}\omega}, (k=1,2,\dots,number\ of\ patches) \quad (20)$$

By finding $V$, the relative displacement of the plate can be defined as:

$$w(x,y,t) = \sum_{i=1}^{N}\sum_{j=1}^{N} U_{ij} W_{ij}(x,y)\left\{\frac{F_0 U_{ij} W_{ij}(x_0,y_0) + \sum_{k=1}^{K}\theta_{ij}A^{-1}[b_1 \cdots b_k \cdots b_n]^T}{\omega_{ij}^2-\omega^2+2j\xi_{ij}\omega_{ij}\omega}\right\}e^{j\omega t} \quad (21)$$

Table 1. Geometric and material properties

| Properties | Plate | Piezoelectric (PZT-5A) |
|---|---|---|
| Young's modulus (GPa) | 70 | 69 |
| Mass density (kg/m3) | 2700 | 7800 |
| Piezoelectric constant (C/m2) | - | -190 |
| Permittivity constant (nF/m) | - | 9.57 |
| Damping ratio | 0.01 | 0.01 |

A. Vibration Metric to evaluate the Shunt Damping Performance

In this study, Equivalent Radiated Power (ERP) is used as the vibration metric to evaluate the shunt damping performance of the system. ERP is mostly used for vibrating panels which includes information about maximum possible acoustic radiation at the excitation frequencies and it is used to demonstrate the amount of reduction on a structure when some sort of vibration control technique is performed [18]. The ERP can be defined as:[1,1,2]

$$\text{ERP}(t) = \frac{1}{2}\rho_f c_f \bigg|_S \dot{w}(x,y,t)^2 dS \quad (22)$$

$$\overline{\text{ERP}(t)} = \frac{1}{2}\rho_f c_f \bigg|_S \dot{\omega}_{rel}(x,y,t)^2 dS \bigg/ \frac{1}{2}\rho_f c_f \bigg|_S dS \quad (23)$$

Once the normalized ERP is calculated, percent reduction compared to the open circuit (OC) condition is calculated using Eq. (24). The results are summarized in Section 5.

$$(Percentage\ reduction\ via\ shunt\ damping)_i = 100 \times \left| \frac{\max(\overline{ERP}_{minimom})_i - \max(\overline{ERP}_{OC})_i}{\max(\overline{ERP}_{OC})_i} \right| \quad (24)$$

### IV. RESULTS AND DISCUSSION

In this section, comparison of the analytical and finite-element models and Electromechanical frequency response functions (FRF) results for connected and separated configurations will be presented. Validation of the analytical model via FEM for a wide range of piezoelectric patch size/thickness is presented in Figure 2 for a single patch. The area ratio is changed from $A_p/A_s = 0.005$ up to $A_p/A_s = 0.75$ (whereas $A_p$ and $A_s$ are the piezoelectric patch and the host structure area, respectively). The thickness ratio is also changed from $h_p/h_s = 0.1$ up to $h_p/h_s = 1.0$

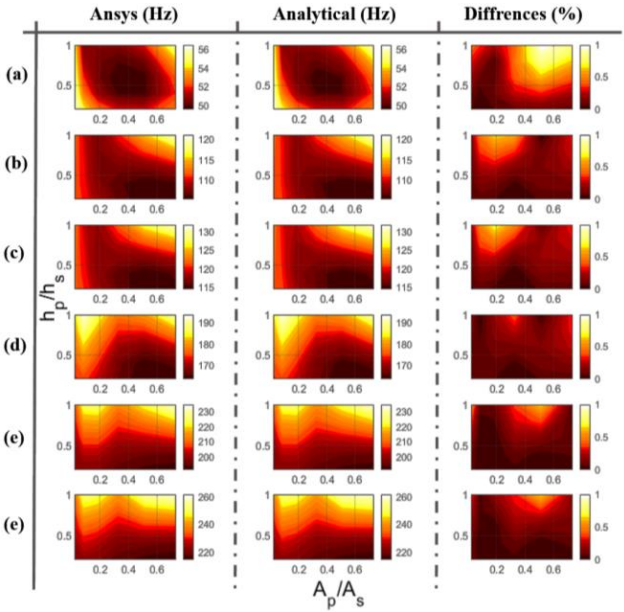

Figure 2: Comparison of the natural frequencies of the first 6 mode shapes (a-e). The first, second and third column shows the Ansys, analytical results, and difference between those results respectively.

Figure 4 shows the comparison of the natural frequencies of the first 6 modes when size and thickness ratios are varied. The first and second column show the Ansys and analytical model results, respectively. The third column shows the percent difference between those results, and it can be observed that for all the first 6 modes of the structure, the difference between the analytical and Ansys model remains within 1%. It shows that the analytical model accurately predicts the natural frequencies of the system when the patch size and thickness are varied.

For a better comparison between cases of separated patches configurations, the $\overline{ERP}$ is investigated for the first 6 modes of the system and 1 patch and 36 patch results are compared in Figure 3. At each step, the is calculated while the R value varied between short circuit (R=100Ω) to open circuit (R=1 MΩ). Then the optimum resistance $R_{opt}$ value is obtained when the $\overline{ERP}$ is minimum over the whole frequency range. The same procedure is repeated for 1 patch and 36 patches. As the numbers of patches are increased, the amplitude of the optimum case is reduced over the full range of frequencies.

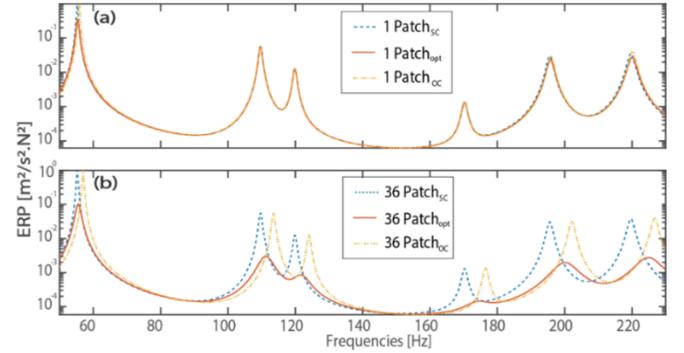

Figure 3: Comparison of $\overline{ERP}$, (a) 1 patch vs (b) 36 patches, dashed blue lines are for the case of SC, solid red lines are for the optimum case and dashed orange ones are for the case of OC.

### V. CONCLUSION

This paper presents a methodology and a formulation for separately shunted piezoelectric patches for achieving higher performance on vibration attenuation. The Rayleigh-Ritz method was used for performing the modal analysis of the electro-mechanical system. The developed model includes mass and stiffness contribution of the piezoelectric patches as well as the electromechanical coupling effect. The vibration performance of the separately shunted piezoelectric patches were compared with the connected configurations. Finally, an optimization study was performed to investigate the effect of the distribution of the resistor values for the first three modes of the structure. The case studies would increase significantly and make the computational cost very high. For that reason, various optimization methods can be explored to reduce the computational cost in future studies.